\documentclass[12pt,preprint]{aastex}

\newcommand{\msun} {$M_\odot$}
\shorttitle{Single Low Mass WDs}
\shortauthors{Kilic, Stanek, \& Pinsonneault}

\begin{document}

\title{The Future is Now: the Formation of Single Low Mass White Dwarfs in the Solar Neighborhood}

\author{Mukremin Kilic, K. Z. Stanek, and M. H. Pinsonneault}

\affil{The Ohio State University, Department of Astronomy, Columbus, OH 43210;\\ kilic, kstanek, pinsono@astronomy.ohio-state.edu}

\begin{abstract}

Low mass helium-core white dwarfs ($M < 0.45$ \msun) can be produced from interacting binary systems, and traditionally all of them have been attributed
to this channel. However, a low mass white dwarf could also result from a single star that experiences severe mass loss on the first ascent
giant branch. A large population of low mass He-core white dwarfs has been discovered in the old metal-rich cluster NGC 6791. There is
therefore a mechanism in clusters to produce low mass white dwarfs without requiring binary star interactions, and we search for evidence
of a similar population in field white dwarfs. We argue that there is a significant
field population (of order half of the detected systems) that arises from old metal rich stars which truncate their evolution prior to the
helium flash from severe mass loss. There is a consistent absence of evidence for nearby companions in a large fraction
of low mass white dwarfs. The number of old metal-rich field dwarfs is also comparable with the apparently single low mass white dwarf population, and our
revised estimate for the space density of low mass white dwarfs produced from binary interactions is also compatible with theoretical expectations.
This indicates that this channel of stellar evolution, hitherto thought hypothetical only, has been in operation in our own Galaxy for many billions of years.
One strong implication of our model is that single low mass white dwarfs should be good targets for planet searches because they are likely
to arise from metal-rich progenitors. We also discuss other
observational tests and implications, including the potential impact on SN Ia rates and the frequency of planetary nebulae.

\end{abstract}

\keywords{stars -- low-mass -- white dwarfs}

\section{Introduction}

The mass distribution of white dwarfs (WDs) in the solar neighborhood peaks at $\sim$0.6 \msun\ (Holberg et al. 2002; Liebert et al. 2005; Kepler et al. 2007).
Separate low mass ($<$0.45 \msun) and high mass ($>$0.8 \msun) components are also observed with each one contributing 10\% and 15\% to the total
mass budget, respectively (Liebert et al. 2005). The magnitude-limited sample of WDs from the Palomar Green (PG) survey includes WDs with masses as low as 0.32 \msun.
Recent discoveries of several extremely low mass ($M\sim$ 0.2 \msun) WDs in the field (Liebert et al. 2004; Eisenstein et al. 2006; Kawka et al. 2006; Kilic et al.
2007a) and around pulsars (Bassa et al. 2006; van Kerkwijk et al. 1996) show that WDs with mass as low as 0.17 \msun\ are formed in
the Galaxy. 

The formation of low mass WDs require significant mass loss from a post main sequence star. The only viable scenario for such mass loss has traditionally been the formation
of a WD in a close binary system. If a post main sequence star has a close binary companion, it can lose its outer envelope without reaching the asymptotic giant
branch and without ever igniting helium, ending up as a helium core WD. Confirmation of the binary nature of several low mass WDs by Marsh et al. (1995),
Marsh 2000, Maxted et al. (2006; 2007), and Kilic et al. (2007b) supports this binary formation scenario.

An alternative scenario for the formation of low mass WDs through extreme
mass loss from single metal-rich stars was proposed by Hansen (2005) to explain
the WD sequence of NGC 6791. Hubble Space Telescope observations of
NGC 6791 showed that it has a population of WDs brighter than expected for a
7 Gyr old open cluster. This population of WDs implied an age of 2.7 Gyr assuming they are carbon-oxygen core stars (Bedin et al. 2005).
Hansen (2005) suggested that the age degeneracy is removed by assuming the stars are He-core WDs and that most of the observed WDs in this cluster are actually low mass WDs
created as a result of significant mass loss on the red giant branch in this metal-rich system ([Fe/H] = +0.3 $-$ +0.5;
Origlia et al. 2006; Gratton et al. 2006).
Follow up spectroscopic work by Kalirai et al. (2007a) showed that the majority
of the brighter WDs in this cluster are actually 0.4 \msun\ WDs. Since the binary star fraction
of NGC 6791 seems to be low (on the order of 14\%; Janes \& Kassas 1997), these low mass WDs are
most likely produced from single star progenitors. 
There is therefore direct observational evidence that low mass WDs may be copiously produced
in a sufficiently old and metal rich star without requiring binary interactions.

In this paper, we provide evidence for the existence of a metal-rich
population of field stars in the past 10 billion years of the Galactic history
and we show that such stars are likely to produce single low mass WDs.
The age-metallicity relation for the local Galactic disk is discussed in \S 2,
while the observed populations of single and binary low mass WDs among the field WD population are discussed in \S 3 and \S 4, respectively.
The implications of our findings are discussed in \S 5.

\section{Metal Rich Stars in the Solar Neighborhood}

The metallicity of the Galactic disk is expected to increase with time as older generations of stars return nucleosynthesis products to the
interstellar environment and new generations are born with more metal-rich gas.  
The age-metallicity relation for FGK dwarfs in the solar neighborhood confirms this expectation, however the presence of old, metal-rich
clusters like NGC 6791 and NGC 188 shows that there is an apparent scatter in age-metallicity relation for the Galactic disk. 

Valenti \& Fischer (2005) presented a large sample of nearby FGK field dwarfs with high resolution spectroscopy which enabled them to determine
stellar properties for these stars including metallicities. Using Hipparcos parallaxes and the above dataset, Reid et al. (2007) created a
volume limited sample of nearby stars within 30 pc of the Sun. They have found that the mean metallicity of nearby stars has increased
by $+0.4$ dex in the last 10 Gyr. 
Figure 1 shows the metallicity histograms for different lookback times for the Reid et al. sample
(kindly made available to us by I. N. Reid). Only stars from the Valenti
\& Fischer (2005) analysis are shown as they have reliable metallicity and age estimates. This figure shows that the mean metallicity of the
nearby sample of FGK dwarfs has increased with time, and that there is a large scatter in metallicity at all ages. Even though the metallicity
distribution peaks below solar metallicity until about 5 Gyrs ago, there have been metal rich stars with [Fe/H] $>$ 0 at all times in the local
Galactic disk. The fraction of metal rich stars increased from 21\% nine billion years ago (middle left panel) to 42\%  five billion years ago
(top right panel). If we consider only stars with [Fe/H] $>$ +0.2 and [Fe/H] $>$ +0.3 (similar to NGC 6791), the fraction of these stars has been
around 14\% and 5\% in the past 10 Gyr, respectively. This analysis shows that NGC 6791 is not unique in terms of its metallicity; $\sim$5\% of the stars
in the solar neighborhood had similar metallicities around the same time NGC 6791 was formed.

Earlier studies by Edvardsson et al. (1993) and Nordstr\"{o}m et al. (2004) 
also found metal-rich stars at all times in the history of the Galactic disk, including stars with [Fe/H] $>$ +0.3 (see the right two panels
in Figure 7 of Reid et al. 2007). Hence, these studies also confirm that metal-rich stars has existed in the Galaxy for a long time.
Since the metal rich stars in NGC 6791 have formed low mass WDs through extreme mass loss in the red giant branch, there is a priori no reason to assume that
the same mechanism would not work for metal-rich field stars. A confirmation of this mechanism would come from the discovery of single low mass WDs in the
field WD sample.

\section{Single Low Mass White Dwarfs}

The binary formation scenario for low mass WDs has been tested several times. The first efforts to find binary companions to low mass WDs
came from Marsh et al. (1995). They have obtained radial velocity measurements of 7 low mass WDs found by Bergeron et al. (1992) and they have
discovered 5 double degenerate close binary systems. Two of the stars in their sample, WD1353+409 and WD1614+136 did not show any radial velocity
variations and their predicted companions still remain to be discovered (Maxted \& Marsh 1998). Based on the high success rate of their observations,
the idea that all low mass WDs has to be produced through close binary evolution became widely accepted. 

A recent analysis of 348 hydrogen atmosphere WDs from the PG survey has revealed 30 low mass WDs with $M<0.45$ \msun\ (Liebert et al. 2005).
Liebert et al. predicted that 100\% of these low mass WDs require binary evolution most likely with other helium or carbon/oxygen core WDs. 
Fifteen of these WDs have already been searched for companions through radial velocity observations or near-infrared photometry.
Eight (53\%) of these low mass WDs have companions found by radial velocity surveys (Marsh 1995; Marsh et al. 1995; Holberg et al. 1995;
Maxted \& Marsh 1999; Nelemans et al. 2005) two of which also confirmed by near-IR photometry (Farihi et al. 2005).
Maxted et al. (2000) did not find any radial velocity companions to 5 PG low mass WDs. In addition, Farihi et al. (2005)
did not see any near-IR excess for 3 WDs in this sample. One object with no evidence of a companion, WD1614+136, is commmon to both Maxted
et al. and Farihi et al. surveys. Overall, 47\% of the PG low mass WDs that have been searched for companions seem to be single.

Maxted et al. (2000) presented their findings from a larger radial velocity survey of 71 WDs including 14 low mass WDs with no detectable radial velocity
variations. They had found 16 low mass WDs with binary companions in their previous surveys, bringing their low mass WD sample size to 30.
This implies that 47\% of the low mass WDs in their extended survey are single. Maxted et al. (2000) found that a mixed single/binary population of
low mass WDs is at least 20 times more likely to explain their data than a pure binary population. Perhaps due to the boldness of the idea at the time,
they concluded that this result depends on several factors including assumed period distribution for possible binary companions, and they found the evidence
for a population of single low mass WDs not sufficient.

The ESO SN Ia Progenitor Survey (SPY) searched for radial velocity variations in more than a thousand WDs using the Very Large Telescope (Napiwotzki et al. 2001).
They have observed about 75\% of the known WDs accessible by the VLT with magnitudes brighter than B=16.5 (Nelemans et al. 2005) including 26 WDs with
$M < 0.42$ \msun. They have found that 15 of these low mass WDs do not show any radial velocity variations, corresponding to a single low mass WD fraction of 58\%
(Napiwotzki et al. 2007). Their detailed analysis of the detection probability of main sequence star, WD, and brown dwarf companions show that
their observations of low mass WDs are inconsistent with the idea that all of them reside in close binary systems.
The SPY observations are less sensitive to brown dwarf companions, because of their smaller mass, but even in this case 100\% binary fraction rate can be ruled out
(R. Napiwotzki 2007, private communication). Brown dwarf companions to WDs are rare (less than 0.5\%; Farihi et al. 2005) and such companions are expected to
end up in short period systems ($P < 1$ day) after the common envelope phase. Such short period brown dwarf companions would reveal their presence through
H$\alpha$ emission as in the case of WD0137-349 (Maxted et al. 2006). 
 
All of the above searches to find companions to low mass WDs show that the fraction of binary low mass WDs is not 100\%. The surveys with more than
10 low mass WDs (the PG, Maxted et al., and SPY surveys) reveal a single low mass WD fraction of 47-58\%.

\section{Binary Low Mass White Dwarfs}

In this section we infer the space densities of
both single and binary helium WDs and compare them with expectations.  We contend that the observed
single He WD population is comparable to what is expected for the fraction of old metal-rich stars.
Furthermore, the binary He WD population is compatible with simple theoretical estimates, and in
particular we argue that the latter population is probably underrepresented because of the presence of
close and bright main sequence (MS) companions.

Liebert et al. (2005) identified three peaks in the WD mass distribution - a low mass component (He) and
two C/O components that correspond to low and high mass evolution. To convert their observations to
formation rates they divided their sample into cooling curve age bins and divided the observed number in each mass bin by
the lifetime in each bin. This provides an observational test of incompleteness, in the sense that if cooler
WDs are missed in the survey, artificially low formation rates would be derived from older and cooler WDs.
This is clearly seen in the most numerous middle category, where the formation rates derived from the three
youngest (and hottest) bins are the same and then decline monotonically for the four oldest bins.
With this procedure they derived formation rates of 0.4, 4.5, and 0.9 $\times 10^{-13} pc^{-3} yr^{-1}$ for the
0.4 \msun, 0.6 \msun, and $>$0.8 \msun\ components, respectively. They also
applied a binary correction to the He WD population.  In the discussion that follows we will omit this because
we are interested in counting systems rather than individual WDs.

In the traditional picture, making a He WD involves truncating the evolution of a low to
intermediate mass star in a close binary system. The maximum mass is set by the size of the core at the end
of main sequence evolution.  Stars above 2.3 \msun\ have core masses large enough upon leaving the main sequence
to produce C/O WDs even for interacting binaries
(for a good review see Iben \& Livio 1993). The maximum orbital separation depends on both
the mass and mass ratio; it can be as small as 0.5 AU for a higher mass  ($\sim$2 \msun) primary with a low mass companion and
as large as 1.5 AU for a lower  mass ($\sim$1 \msun) primary with a higher mass companion.  We adopt 1 AU as an
average value.  For the Duquennoy \& Mayor (1991) period distribution, 12\% of binaries have orbital
periods less than 1 AU.  For a 50\% binary population, this yields a predicted frequency of 6\% for
low mass systems with at least one He WD.

If we exclude the high mass WDs (as close binaries in this mass range
would not necessarily produce He WDs; see Nelemans et al. 2001) the inferred formation rate
of single He WDs in the PG survey is 4\% (assuming 50\% single WD fraction as implied by the surveys discussed in the previous section)
and that of binary He WDs is also 4\%.  Given both the roughness of the calculation and
the small He WD sample in Liebert et al. (2005), the binary fraction is consistent with expectations.  However, there is
a potential - and checkable - systematic error that will lead to a differential undercounting of even hot He WDs from the
presence of MS companions.  
For a typical distance of 100 pc and a Duquennoy \& Mayor (1991) period distribution, one would
expect 50\% of C/O WDs to be single, 20\% to be in resolved binaries, and 30\% to be in unresolved binaries.
By their nature, binary He WDs come from a population that consists entirely of unresolved binaries.
Furthermore, the secondaries of close binaries with periods less than one year
are systematically higher in mass than those of distant binaries; as a result the unresolved main-sequence (MS) companions
of He WDs will be substantially brighter than those of C/O WDs.  Therefore, the detectability of even hot He WDs can
be strongly influenced by companions, and we argue that the Liebert et al. estimate is probably a lower bound
to the number of He WDs from binaries.

The median temperature for the 30 low mass WDs in the PG survey is $T_{\rm eff}\sim25000$ K corresponding to $M_B\sim9.4$ for a 0.4 \msun\ WD (Bergeron et al. 1995).
This is comparable to the blue magnitude for a 0.65 \msun\ MS star; we adopt this as the mass threshold for secondaries that could influence the
detectability of WDs at this temperature.  If we adopt a typical age of 5 Gyr, secondaries with masses below 1.3 \msun\ 
will on average be unevolved.  The ratio of double degenerate systems to those with MS companions has been
addressed in detailed population synthesis studies (Iben et al. 1997).
Iben et al. assumed that only WDs with companions less than 0.3 \msun\ were detectable, and such low
mass systems would be 37\% of the detectable total.  For a flat relative mass function this implies that
in the underlying population 70\% of all systems would have MS companions with masses less than 1.3 \msun.
However, for hotter WDs one would detect a larger fraction of systems. Out of all systems 32\% would be undetected ($M_{\rm secondary} >$ 0.65 \msun),
38\% detected with an MS companion ($M_{\rm secondary} <$ 0.65\msun), and 30\% would be double degenerates. With this assumption an observed relative formation
rate of 4\% for binary WDs becomes an intrinsic population of 5.9\%. This is entirely consistent with the predicted binary fraction of 6\% from
Duquennoy \& Mayor (1991) period distribution.

We can use the 2MASS data to check on the fraction of systems with low mass companions.
With a field binary fraction of 50\% we would expect half of the ``single'' WDs to actually have distant and unresolved
non-interacting binary companions; these companions will typically be low mass stars. We expect 70\% of the low mass WDs produced from
binary interactions to have close MS companions and the other 30\% to be double degenerate systems. In the Liebert et al. sample there
are 14 low mass WDs with $T_{\rm eff} > 25,000$ K that were used to estimate the formation rate without applying an incompleteness correction. We
would expect 2 double degenerate systems, 5 close binary companions, and 3.5 distant binary companions. Three systems are known to have
degenerate companions, and we find significant J and K band excesses for 4 of the systems. Based upon a comparison with isochrones, we
infer masses of 0.3-0.35 \msun\ for 3 companions and $\sim$0.5 \msun\ for one. This indicates a deficit of companions relative to expectations, and
in particular we note that there are no higher mass companions seen even though a significant number would be naturally produced by binary
evolution. If all systems were from interacting binaries, and the sample was complete, we would expect 10 MS companions. We therefore
conclude that even the hot sample of He WDs is likely to be incomplete due to the presence of nearby companions. Until a radial velocity
survey is completed, it is also not clear whether the MS companions detected are close binaries (attributed to the binary formation channel)
or distant ones (attributed to the single formation channel). Follow-up work of this type will be essential to empirically set the
intrinsic space densities.

\section{Discussion}

More metal-rich post-main sequence stars are expected to lose more mass on the red-giant branch compared to less metal-rich stars, and this effect is
directly observed in the red-giant branch luminosity functions of several old open clusters including NGC 188 and NGC 6791 (see Fig. 8 in Kalirai et al. 2007a).
An important question is the significance of the mass loss process for different metallicities. Kalirai et al. (2005; 2007a) claimed to see a small dependence
of mass loss on metallicity in Hyades ([Fe/H] = +0.17) and NGC 2099 (M37; [Fe/H] = $-$0.1).

For a given fraction of low mass WDs produced from severe giant branch mass loss, one can compare with the field metallicity distribution to
infer a critical abundance above which the late stages of stellar evolution are truncated. Our implied assumption is that all old stars
above some threshold truncate their evolution above a critical metallicity and none do below it; this is certainly a simplification given
the stochastic nature of mass loss. We also note that this process is expected to occur only for stars that are old enough (and have low
enough turnoff mass) to shed their envelopes prior to experiencing a helium flash. An intermediate mass metal-rich star would still
produce an ordinary C/O WD even with enhanced mass loss. As a result, the boundary that we derive is an upper limit, and the
process may become important for stars with even lower metal abundances.

According to the Reid et al. (2007) analysis, the fraction of [Fe/H] $>$ 0 stars in the solar neighborhood has been more than 21\% in the past 10 Gyrs.
If the mass loss process was
significant for [Fe/H] $>$ 0, we would expect to find 21\% of the field WD population as single low mass WDs. Since the single low mass WD formation rate among
all WDs is around 4\%, the mass loss is not significant enough for [Fe/H]$>$ 0 to produce WDs with helium cores ($<$0.45 \msun). 
The fraction of metal rich stars with [Fe/H] $>$ +0.2 has been more than 10\% in the past 10 Gyr, therefore the metallicity effect seems to kick in for more
metal-rich stars. We find that the mass loss becomes significant enough to produce low mass WDs for [Fe/H] $>$ +0.3 as the fraction of [Fe/H] $>$ +0.3 stars
has been $\geq$3\% for the past 10 Gyr. 

A caveat in this analysis is that the Valenti-Fischer dataset is biased toward more metal rich stars than the complete volume-limited sample of Reid et al. (2007;
see their Figure 6). Therefore, the actual metallicity limit where the mass loss process is extreme enough to create low mass WDs may be lower than [Fe/H]  = +0.3.
For example, Allende Prieto et al. (2004) found that among the 1477 thin disk stars in their extended sample, the fraction of stars with [Fe/H] $>$ +0.2 and +0.3
is 5.6\% and 1.8\%, respectively. 

Observations of several star clusters with different metallicities can help refine the constraint on metallicity at which the mass loss is significant enough to
create low mass WDs, however the field population of metal rich stars and single low mass WDs imply that low mass stars with [Fe/H] $\geq$ +0.3 are likely to lose enough
mass on the red giant branch to skip the asymptotic giant branch evolution and directly evolve into single helium-core WDs.

The confirmation of the mass loss - metallicity dependence for both NGC 6791 and the field stars has important implications: 

\subsection{Planets Around WDs}

Even though several groups have been searching for planets around WDs through infrared (Debes \& Sigurdsson 2002 ; Burleigh et al. 2002) and mid-infrared
excess around WDs (Mullally et al. 2007), there are no known planetary companions to WDs yet. This may be caused by a target selection effect.
As all of these groups have targeted typical $\sim$0.6 \msun\ WDs, they have selectively targeted descendants of lower metallicity stars.
Fischer \& Valenti (2005) found that at [Fe/H] $>$ +0.3, 25\% of the stars have gas giant planets whereas less than 3\% of the stars with -0.5 $<$ [Fe/H] $<$ 0.0
have Doppler-detected planets. Since the super-solar metallicity stars are expected to be the progenitors of single low mass WDs, these WDs are prime targets
for planet detection. If the progenitors of these single low mass WDs have planets (25\% likely), the planets are more likely to survive the post-main sequence
evolution, as these stars do not go through the AGB phase.

\subsection{Extremely Low Mass WDs}

Even though the fraction of binary low mass WDs seems to be around 50\% for $\sim$0.4 \msun\ WDs, we expect this fraction
to be 100\% for the extremely low mass WDs with $M\leq0.2$ \msun. The mass loss process is most likely not significant enough to produce 0.2 \msun\ WDs
even at higher metallicities. The initial-final mass relation for WDs (Weidemann 2000; Kalirai et al. 2007b) predict that a 1 \msun\ star would produce a 0.55 \msun\ WD. On
the other hand, at [Fe/H] $\sim$ +0.4, D'Cruz et al. (1996) predict that the same star would produce a 0.44$-$0.47 \msun\ WD, well above the 0.2\msun\ limit.
So far most of the known extremely low mass WDs are found as companions to pulsars. Only recently, several 0.2 \msun\ WDs are found in the field population
(Liebert et al. 2004; Kawka et al. 2006; Eisenstein et al. 2006; Kilic et al. 2007a). However, only one of them has been searched for companions and found
to be in a binary system (Kilic et al. 2007b). A systematic search for companions to WDs with masses 0.2$-$0.4 \msun\ is required to evaluate the binary fraction
at different mass ranges.

\subsection{Red Giants}

A prediction of this scenario is that we should see very few truly metal rich red clump giants. This effect is seen in the red giant branch
luminosity function of NGC 6791; the stars are thinning out of the red giant branch and some stars never reach the tip of the giant branch (Kalirai et al. 2007a).
A similar effect is also seen in the local population of giants. Figure 9 in Luck \& Heiter (2007) compares the metallicity histograms for dwarfs and giants
within 15 pc of the Sun. It is clear from their figure that even though the dwarf population has a high metallicity tail extending up to [Fe/H]$\sim$+0.6
(for both of their dwarf samples including and excluding the planet-hosting stars; top three panels), the giants show a significant drop in numbers after
[Fe/H] = +0.2 and no giants with [Fe/H] $>$ +0.45 are observed in the field population. 

Using Hipparcos parallaxes and metallicities for 284 nearby red giant stars, Udalski (2000) showed that 
there is a very well defined red clump for lower metallicity giants, i.e. [Fe/H] $< -$0.05, but above that the red clump becomes
really broad (see their Figure 2). The higher metallicity bin is poorly defined due to the small number of metal-rich red clump stars
in the solar neighborhood. Depending on the metallicity of the progenitors, the mass loss process is likely to create red giants with a
range of envelope masses which would create a broad distribution of red clump stars. These observations also suggest that the mass loss can
thin out the envelopes of red giants and that most super-solar metallicity stars will not reach the tip of the red giant branch.

\subsection{Type Ia SNe} 

In our scenario, the mass loss rate in cool giants increases dramatically with increased metallicity. This could impact the winds
from more massive stars as well, which in turn could affect the Type Ia SNe production rate. 
There are two main formation scenarios for Type Ia SNe; accretion from a non-degenerate companion star onto a WD close to the Chandrasekhar
mass limit (e.g. Whelan \& Iben 1973) and mergers of two WDs (e.g. Webbink 1984). In the former case, a finely tuned mass accretion rate is required to
avoid either nova explosions or common envelope evolution. Changes in the stellar wind properties will certainly impact this balance. 
A low SNIa production rate is actually observed in elliptical galaxies (e.g. Sullivan et al. 2006) and in low-redshift galaxy clusters (Sharon et al. 2007)
relative to star forming galaxies. This has been interpreted as evidence that a prompt channel for SNIa is more important than the slower
phase which had previously been believed to be the main channel (e.g. Panagia et al. 2007). However, a metallicity dependent production efficiency could explain the
same observations and should be explored. 

\subsection{PNe}

Our understanding of the formation of planetary nebulae (PNe) and whether they all require binary companions or not is questionable (Moe \& De Marco 2006).
If both binary and single stars produce PNe, another interesting consequence of our finding is that these low-mass,
helium core WDs would most likely not produce PNe, as they
completely avoid the AGB phase. Indeed, Gesicki \& Zijlstra (2007) find that
the central stars of PNe have a very sharply peaked mass distribution, with a mean mass of $M=0.61\; M_{\odot}$
and a range of $0.55-0.66$ \msun. The relatively low number of PNe (per unit galaxy luminosity)
in more metal-rich elliptical galaxies (Buzzoni et al. 2006; see their Figure 11) is consistent with this prediction.
 
\section{Conclusions}

We examined the age-metallicity relation for the local Galactic disk and found that there have been super-solar metallicity stars in the solar neighborhood
in the past 10 Gyr. Since high metallicity stars in NGC 6791 produced single low mass WDs, we expect that there should be a population of single low mass WDs
in the field WD population as well. Such a population of single low mass WDs is actually observed in several surveys including the Palomar Green sample and
the SPY survey. The observed fraction of single low mass WDs is around 50\%, corresponding to a formation rate of 4\% among the entire WD population.
By comparing the fraction of high metallicity stars in the solar neighborhood, we show that $\geq$3\% of the stars in the solar neighborhood have had
[Fe/H] $>$ +0.3 in the past 10 Gyr, therefore such stars are likely progenitors of the single low mass WDs observed today.
Our analysis indicates that this channel of stellar evolution actually happens not only in NGC 6791, but in the field stars as well.
The major implication of our finding is that a substantial fraction of old metal-rich stars end their lives prior to igniting helium and
skip interesting and visible phases of stellar evolution that such stars would ordinarily experience. This also leads to an inversion of
prior searches for planetary companions to WDs, which avoided low mass WDs as being the products of binary interactions.
Instead, some of these objects may be the descendants of the sort of metal-rich stars that are known to have a high specific frequency of
planet detection.

\acknowledgements
We would like to thank I. N. Reid for a careful reading of this manuscript. We would also like to thank the participants of the morning ``Astronomy Coffee'' at
the Department of Astronomy, The Ohio State University, for the daily and lively astro-ph discussion, one of which prompted us to investigate the problem
described in this paper.

\clearpage
\begin{figure}
\plotone{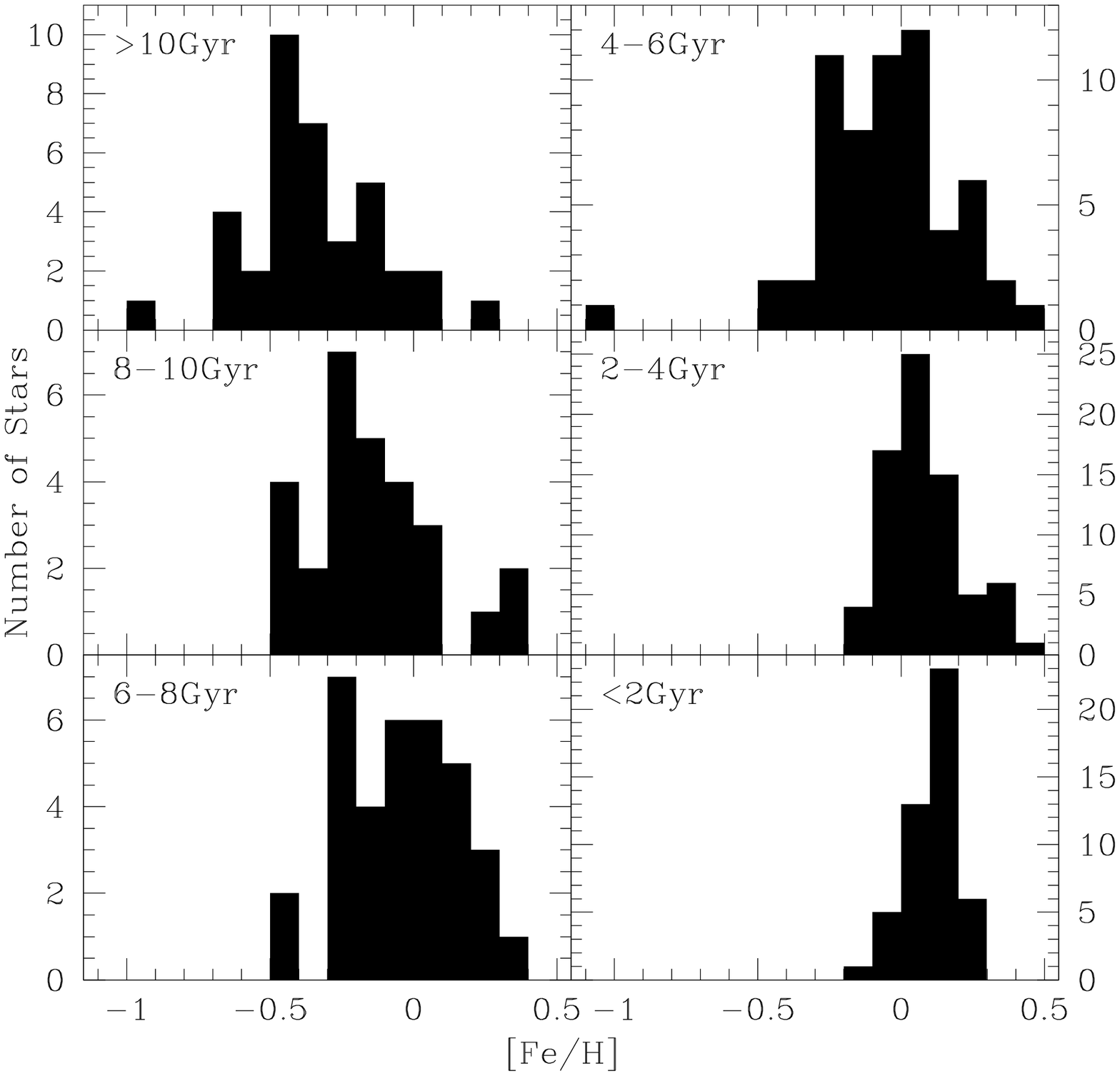}                
\caption{Metallicity histograms for the 30 pc volume limited sample of nearby stars from Reid et al. (2007). Only stars with metallicity
measurements from the high resolution spectroscopic analysis of Valenti \& Fischer (2005) are shown.
Metal enrichment of the local Galactic disk with time is apparent in this figure. The presence of metal rich stars at all
times shows that stars like these are likely progenitors of single low mass WDs observed in the nearby WD sample.}
\end{figure}

\end{document}